\documentstyle[epsf,aps,prl]{revtex}

\begin{document}
\draft

\twocolumn[
\hsize\textwidth\columnwidth\hsize\csname@twocolumnfalse\endcsname

\title{Stiction, Adhesion Energy and the Casimir Effect in Micromechanical Systems } 

\author{E. Buks and M. L. Roukes}

\address{Condensed Matter Physics, California Institute of Technology, Pasadena, CA 91125} 

\date{\today} 
 
\maketitle 
 
\begin{abstract} 

We measure the adhesion energy of gold using a micromachined doubly-clamped beam.  The stress and stiffness of the beam are 
characterized by measuring the spectrum of mechanical vibrations and the deflection due to an external force.  To determine 
the adhesion energy we induce stiction between the beam and a nearby surface by capillary forces.  Subsequent analysis
yields a value $\gamma =0.06$ J/m$^{2}$ that is a factor of approximately six smaller than predicted by idealized 
theory.  This discrepancy may be resolved with revised models that include surface roughness and the effect of adsorbed 
monolayers intervening between the contacting surfaces in these mesoscopic structures.

\end{abstract} 

]

The Casimir effect \cite{793} is a striking consequence of quantum
electrodynamics (for a recent review see \cite{850}). \ The dependence of
the ground state energy of the electromagnetic field upon boundary
conditions gives rise to an observable force between macroscopic bodies. \ A
significant enhancement in the accuracy of measuring the Casimir force has
been achieved recently with experiments employing torsion pendulum \cite{5}
and atomic force microscope (AFM) \cite{4549}. \ Casimir effect
investigations may open the way for experimental observation of new
fundamental forces arising from the hypothetical extra dimensions predicted
by modern theories \cite{23}. \ However to enable such studies it is crucial
to improve experimental techniques. \ The Casimir force, in addition to its
fundamental interest, also plays an important role in the fabrication and
operation of micro electromechanical systems (MEMS). \ This technology
allows fabrication of variety of on-chip fully integrated sensors and
actuators with rapidly growing number of applications. \ One of the
principal causes of malfunctioning in MEMS is {\it stiction}, namely
collapse of movable elements into nearby surfaces, resulting in their
permanent adhesion (for a review see \cite{385}, \cite{1}). \ This can occur
during fabrication, especially due to capillary forces present during drying
of a liquid from the surface of the sample, or during operation. \ It was
argued recently that the Casimir effect is often an important underlying
mechanism causing this phenomena \cite{2501}.

Here we report our experimental study of surface - surface interactions
using micromachined, doubly-clamped Au beams. \ In particular, we focus upon
the extreme manifestation of the Casimir interaction, namely, adhesion
between surfaces and the associated energy of this process. \ The structures
we use are designed to allow straightforward and unambiguous interpretation
of our results. \ We use bulk (rather than surface) micromachining, in which
the substrate is completely removed beneath the sample. \ This greatly
simplifies the boundary conditions of the electromagnetic field in the
vicinity of the sample. \ Moreover, we avoid using multilayered structures,
since their internal stresses generally play an important role and
theoretical modeling is thus more difficult. \ We use metallic rather than
semiconductor structure to minimize the possibility of parasitic attractive
forces arising from bound surface charge.

After characterizing the mechanical properties of the beam, we induce
stiction between the beam and a nearby electrode. \ The shape of the beam
after adhesion and the elastic energy associated with this configuration
allow us to determine the attractive surface energy. \ Similar methods were
employed to measure the adhesion energy of stress-free Si \cite{385}, \cite
{1}. \ For such work it is important that mechanical properties such as
stress be well characterized to obtain an accurate determination of the
elastic energy. \ We conclude by comparing our results with previous
measurements and with theory.

The bulk micromachining process employed for sample fabrication is described
in Fig. 1. \ In the first step chemical vapor deposition is employed to
deposit 70 $%
\mathop{\rm nm}%
$ thick layer of low-stress silicon nitride on the front and back sides of a
Si wafer. \ A square window is then opened in the back side layer using
photolithography (Fig. 1(a)). \ An anisotropic and selective wet etching using KOH forms the structure seen in Fig. 1(b). \ The gold beam and its adjacent electrodes are patterned on the front side of the
resulting 300$\mu $m square membrane of silicon nitride (Fig. 1(b)). The nanofabrication steps
for this part of the process involve electron beam (e-beam) lithography and
thermal evaporation (Fig. 1(c)). \ \ The complete beams have length $l=200$ $%
\mu $m, width $a=0.24\mu $m and thickness $t=0.25\mu $m (measured using
AFM). \ In the final step the membrane is removed using electron cyclotron
resonance plasma etching with an Ar/NF$_{3}$ gas mixture bombarding the back
side of the sample. This leaves the gold beam suspended (Fig. 1(d)). \
Figure 1(e) is a micrograph showing a side view of the device.

To characterize the mechanical properties of the beam we employ two methods,
namely, measurement of the resonance frequencies of the beam and measurement
of the deflection due to an external force. \ Both methods lead to similar
conclusions.

The equation of motion of the beam is given by:

\begin{equation}
\frac{\partial ^{2}y}{\partial x^{2}}-\zeta ^{2}l^{2}\frac{\partial ^{4}y}{%
\partial x^{4}}=\left( \rho A/T\right) \frac{\partial ^{2}y}{\partial t^{2}}%
-f\,/T,  \label{eom}
\end{equation}
where \smallskip $\zeta ^{2}=EAa^{2}/12Tl^{2},$ with $E$ being Young's
modulus, $A=at$ is the area of the beam's cross section, $T$ is the tension, 
$\rho $ is the mass density, and $f$ is the density of external force \cite
{elas}. \ The clamping of the beam on both sides is taken into account using
the boundary conditions $y\left( \pm l/2\right) =\frac{\partial y}{\partial x%
}\left( \pm l/2\right) =0$.

The dimensionless parameter $\zeta $ indicates the relative effect of
stiffness compared with tension on the dynamics of the beam. \ As we shall
see below, $\zeta <<1$ in our case, therefore we expand the resonance
frequencies of the system in powers of $\zeta $ using perturbation theory. \
To second order we find:

\begin{equation}
\nu _{n}=n\nu _{0}\left[ 1+2\zeta +\left( 4+n^{2}\pi ^{2}/2\right) \zeta ^{2}%
\right] ,  \label{fre}
\end{equation}
where $\nu _{0}=\sqrt{T/\rho A}/2l$. \ The equally spaced spectrum obtained
for the case $\zeta =0$\ is the same as for a stiffness free beam with
boundary conditions $y\left( \pm l/2\right) =0$. \ Note that the terms that
make the spectrum unequally spaced are of order $O\left( \zeta ^{2}\right) .$

The resonance frequencies are measured {\it in-situ} using a commercial
scanning electron microscope (SEM). \ The electron beam is focused on a
point near the edge of the gold beam and the output signal from a
photomultiplier (serving as a secondary electron detector) is monitored
using a spectrum analyzer to detect mechanical displacement (see Fig. 2(a)).
\ Note that this detection scheme is sensitive almost exclusively to motion
in the plane of the sample.

Without applying any external excitation we find a pronounced peak near $\nu
_{1}=176.5%
\mathop{\rm kHz}%
$ associated with thermal excitation of the fundamental mode of the beam
(see Fig. 2(b)). \ The thermal peaks of higher modes are too small to be
detected, therefore we induce external excitation by applying an AC voltage
to a nearby parallel electrode, separated from the beam by a gap of width $%
g=5\mu $m. \ We find three higher modes with frequencies $\nu _{2}=354.4%
\mathop{\rm kHz}%
$, $\nu _{3}=529.8%
\mathop{\rm kHz}%
$, and $\nu _{4}=709.7%
\mathop{\rm kHz}%
$. \ The fact that the obtained spectrum is almost equally spaced indicates
that $\zeta <<1$. \ Note, however, that drift in the position of the peaks
occurring over time prevents us from making a precise estimation of $\zeta $%
. \ Based on the uncertainty originated by this drift we place an upper
bound of $\zeta <0.015$.

Theoretically, the spectral density of displacement noise near the center of
the beam ($x=l/2$) around the fundamental frequency for the case $\zeta =0$
is given by:

\begin{equation}
S_{x}\left( \omega \right) =\frac{\omega _{0}k_{B}\Theta }{\pi Qm_{\text{eff}%
}\left[ \left( \omega _{0}^{2}-\omega ^{2}\right) ^{2}+\left( \omega
_{0}\omega /Q\right) ^{2}\right] },  \label{peak}
\end{equation}
where $Q$ is the quality factor, $m_{\text{eff}}=\rho Al/2$ is the effective
mass, $\omega =2\pi \nu $ is angular frequency, and $\Theta $ is the
temperature. \ Fitting the data in Fig. 2(b) with Eq. (\ref{peak}) yields $%
Q=1800$. \ The known parameters of the beam allow determination of the
scaling factor translating the signal of the spectrum analyzer to actual
displacement noise. \ Using this factor and the signal to noise ratio of the
data in Fig. 2(b) we find the sensitivity of our displacement detection
scheme to be $4\times 10^{-13}%
\mathop{\rm m}%
/\sqrt{%
\mathop{\rm Hz}%
}$. \ This value can be further enhanced by increasing the current of the
electron beam. \ However, to minimize heating of the device due to electron
bombardment we operate at a relatively low current of 100 pA. \ The energy
absorbed by the sample depends on the penetration depth of electrons and on
the thickness of the Au layer. \ For an acceleration voltage of 40 kV we
estimate the heating power is of order 100 nW \cite{scanning}. \ For thermal
conductivity of 300 W/mK and the geometry of our device the temperature
increase is $\approx $1K.

To further establish our findings we study the deflection of the beam due to
application of a uniform force. \ For this we apply D.C. voltage $V$ between
the beam and the nearby electrode. \ When the deflection is small compared
to the distance between the beam and the electrode the force acting on the
beam is approximately uniform. \ The expected deflection is found from the
steady state solution of Eq. (\ref{eom}) with $f=$constant:

\begin{equation}
y\left( x\right) =\frac{fl^{2}}{2T}\left[ \frac{1-\left( 2x/l\right) ^{2}}{4}%
+\frac{\zeta \left[ \cosh \left( x/\zeta l\right) -\cosh \left( 1/2\zeta
\right) \right] }{\sinh \left( 1/2\zeta \right) }\right] .  \label{def}
\end{equation}

Figure 3(a) shows a series of SEM pictures taken with $V=$0, 10,..., 70 V. \
Using image processing we extract the shape of the beam in each picture,
namely the experimental value of $y\left( x\right) .$ \ Comparing the
calculated $y\left( x\right) $ with experimental data using a least squared
fit, we determine the parameter $\zeta =0.014\pm 0.007$, in agreement with
the above mentioned estimate of $\zeta $.

The value $\zeta =0.01$ and the other known parameters allow estimating of
Young's modulus $E=8\times 10^{10}$ N/m$^{2}$. \ This value shows reasonable
agreement with previous measurements of $E$ in thin films of evaporated gold
using different methods \cite{1096}, \cite{229}, \cite{931}.

Figure 3(b) shows the maximum displacement of the beam, namely $y\left(
0\right) $, as a function of the voltage $V$. \ As expected, we find that
this maximum displacement is proportional to $V^{2}$. \ Using the value of $%
T=5.8\times 10^{-6}%
\mathop{\rm N}%
$ found from the spectrum measurements we find that $f/V^{2}=4.6\times
10^{-7}$ N/mV$^{2}$

To study adhesion in our system we bring the beam and the nearby electrode
to contact by introducing a pure liquid to the surface of the sample and
employing the resultant capillary forces. \ During drying a thin layer of
liquid is formed between the gold surfaces. \ The pressure inside the drop
is lower than the pressure outside if the wetting angle is smaller than $\pi
/2$, resulting in a net attractive force between the surfaces. \ We employ
DI water as an adhesive liquid due to its relatively high surface tension ($%
\cong 0.07$ N/m at room temperature).

Fig. 4 is a micrograph of the gold beam after drying the DI water from the
surface of the sample. \ The length of the segment that adheres is $%
d=67.8\mu 
\mathop{\rm m}%
$. \ The fact that adhesion between the beam and the nearby electrode
persists after drying indicates that the total energy of the adhering system
is lower than that of a straight free beam, which is merely metastable.

To estimate the total energy of the system we make two simplifying
assumptions: (a) no stiffness, namely $\zeta =0$ (the measured value $\zeta
\simeq 0.01$ justifies this approximation) ; (b) no finite range interaction
between the surfaces (the error due to this approximation is small due to
the rapid decay of the interaction as a function of distance). \ Using the
first assumption we find an expression for the elastic energy of the system

\begin{equation}
U_{e}=2g^{2}T/\left( l-d\right) .  \label{elas}
\end{equation}

The second assumption implies that the energy due to the surface-surface
interaction is given by:

\begin{equation}
U_{a}=-st\gamma ,  \label{adh}
\end{equation}
where $\gamma $ is the energy of adhesion per unit area. \ The condition
that the total energy of the system has a minimum implies:

\begin{equation}
\gamma =2g^{2}T/t\left( l-d\right) ^{2}.  \label{gamma}
\end{equation}

Using the parameters of our sample we find $\gamma =0.066%
\mathop{\rm J}%
/%
\mathop{\rm m}%
^{2}.$ \ A similar value of $0.062%
\mathop{\rm J}%
/%
\mathop{\rm m}%
^{2}$ is obtained from another beam with a gap $g=3$ $\mu $m$.$

What is expected theoretically? \ The Casimir force for small separation is
reduced to the so-called non retarded van der Waals force \cite{israel}. \
In this regime the separation between the metallic surfaces is small
compared to the characteristic wavelength of their absorption spectra and
effects due to finite conductivity are strong. \ The interaction energy per
unit area is given by

\begin{equation}
U=-A/12\pi d^{2}.  \label{vdw}
\end{equation}
For the case of Au it was found that Eq. (\ref{vdw}) is a good approximation
for $d<2$ nm and the Hamaker constant $A$ is given by $A=4.4\times 10^{-19} %
\mathop{\rm J} $ \cite{3093}. \ This allows estimation of the adhesion
energy by $\gamma =A/12\pi d_{0}^{2}$, where $d_{0}$ is the effective
separation at contact. \ The nearest neighbor approximation for the case of
atomically flat surfaces leads to $d_{0}\approx 0.16%
\mathop{\rm nm}%
$\cite{israel}, and therfore $\gamma \approx 0.4%
\mathop{\rm J}%
/%
\mathop{\rm m}%
^{2}$. \ For the case of metals it was shown that electron exchange
interaction (giving rise to the so-called metallic bond) is expected to
further enhance $\gamma $ \cite{3427}. \ The enhancement factor, however,
strongly depends on the twist angle between the contacting lattices. \
Previous measurements of $\gamma $ of metals found values in the range of $%
0.4-4%
\mathop{\rm J}%
/%
\mathop{\rm m}%
^{2}$ \cite{israel}.

There are two possible explanation for the factor of six discrepancy between
our results and theory. \ The first is roughness existing on the surfaces in
contact. \ From the measured value of $\gamma $, the calculated value of $A$%
, and the relation $\gamma =A/12\pi d_{0}^{2}$ we find an effective value
for the separation between the surfaces within our sample of $d_{0}\approx
0.4%
\mathop{\rm nm}%
$. \ Note that this distance scale for $d_{0}$ is far smaller than can be
resolved using SEM or AFM. \ Another possible cause for the discrepancy
might be surface contamination that can strongly modify the adhesion energy
even when this is from absorbed layers only a monolayer thick \cite{israel}.

Apart from determining the adhesion energy, a central question is whether we
can study the Casimir interaction at finite separation with such stiction
experiments. \ In the immediate vicinity of the region of the beam that is
in contact with the electrode, the separation between the beam and electrode
is quite small. This gives rise to a strong Casimir interaction in this
locale. \ In principle such attraction can cause additional bending of the
beam, thereby allowing determination of the magnitude of the attractive
force using Eq. (\ref{eom}
) . \ To explore this possibility we estimate this additional bending
assuming the attractive interaction is given by Eq. (\ref{vdw}) with $%
A=4.4\times 10^{-19}%
\mathop{\rm J}%
$. \ We assume $\zeta =0$ and solve Eq. (\ref{eom}) using the other known
parameters of the beam. \ We find that the change in the separation between
the beam and the electrode becomes comparable to the unperturbed one only
when the separation is $<1%
\mathop{\rm nm}%
$. \ Resolving such a small effect is very difficult with a SEM but might be
possible by imaging with a transmission electron microscope (TEM), if sample
charging is carefully controlled. \ However, we find that the effect of
stiffness on the shape of the beam is much stronger than the effect due to
Casimir attraction. \ We note that observation of Casimir-induced bending
may be easier using a stress free material with a low Young's modulus and
employing a modified geometry.

As demonstrated by the present work, MEMS can provide ideal tools for
characterizing stress in thin films, as well as for studying adhesion
forces. \ Future experiments with enhanced sensitivity should enable studies
of the Casimir force at finite separations.

The authors are grateful to K. Schwab for his assistance in sample
fabrication and Y. Buks for image processing of the SEM micrographs. \ This
research was supported by DARPA MTO/MEMS under grant DABT63-98-1-0012. \
E.B. gratefully acknowledges support from a Rothschild Fellowship, and from
an R. A. Millikan Fellowship at Caltech.

\newpage 
\begin{figure}[b]
\caption{The device is fabricated using bulk micromachining techniques. \ In
steps (a) and (b) a suspended membrane of silicon nitride is formed. \ A
gold beam is fabricated on top of the membrane (c) and the membrane is
etched, leaving the beam suspended (d). \ Side view micrograph of the device
is seen in (e).}
\label{fig:fig1}
\end{figure}

\begin{figure}[b]
\caption{(a) The setup employed to detect the resonance frequencies of the
beam. \ (b) peak in the displacement noise associated with thermal
excitation of the fundamental mode.}
\label{fig:fig2}
\end{figure}

\begin{figure}[b]
\caption{(a) deflection of the beam due to application of electrostatic
force. \ (b) Displacement of the center of the beam as a function of applied
voltage.}
\label{fig:fig3}
\end{figure}

\begin{figure}[b]
\caption{Adhesion between the beam and a nearby electrode.}
\label{fig:fig4}
\end{figure}

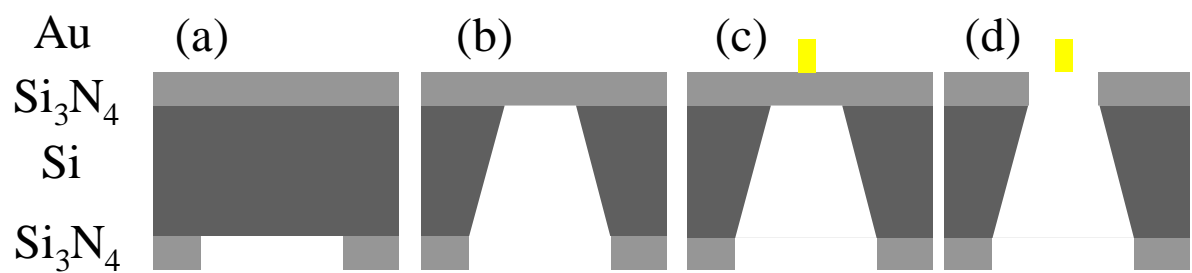
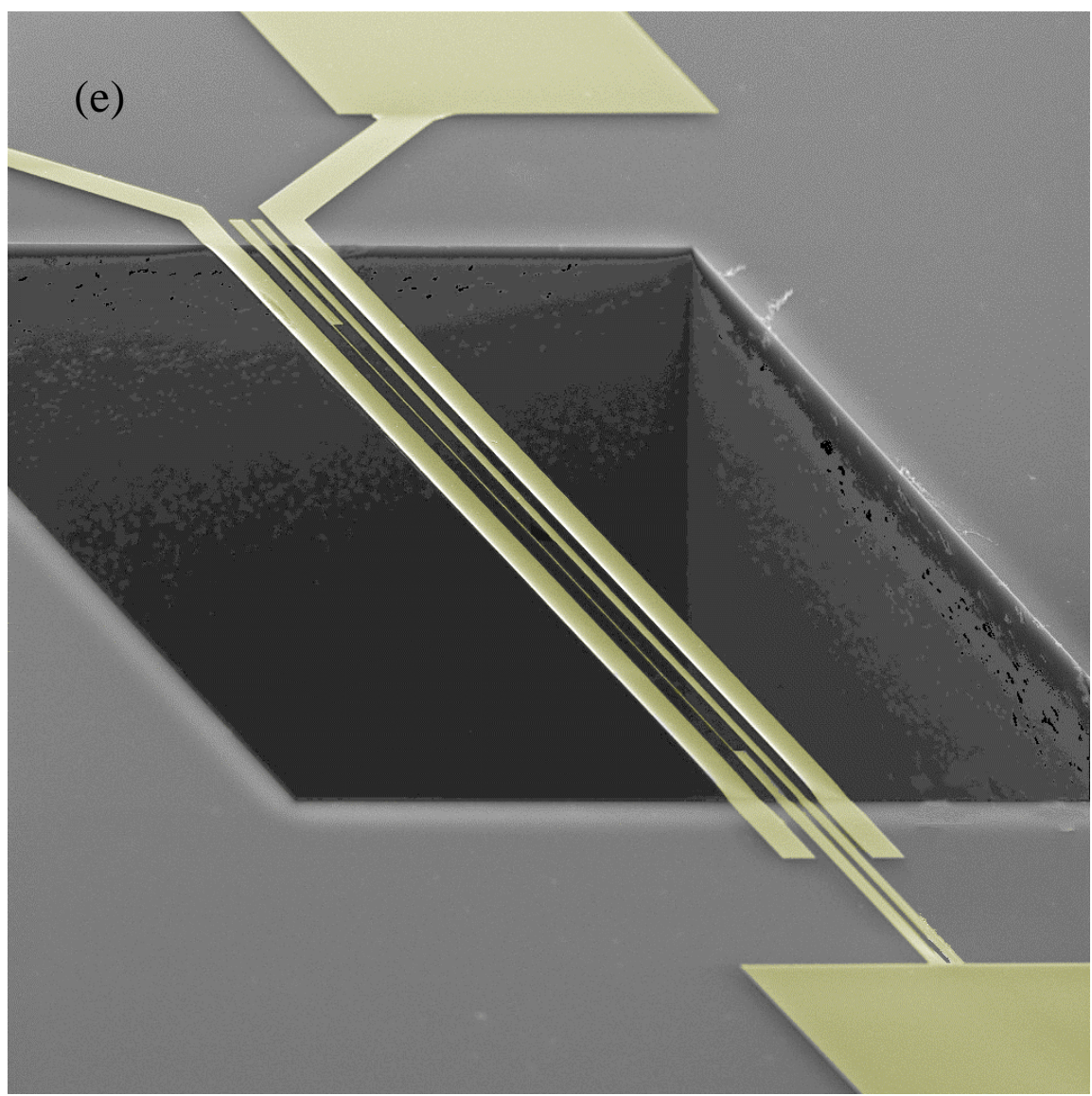
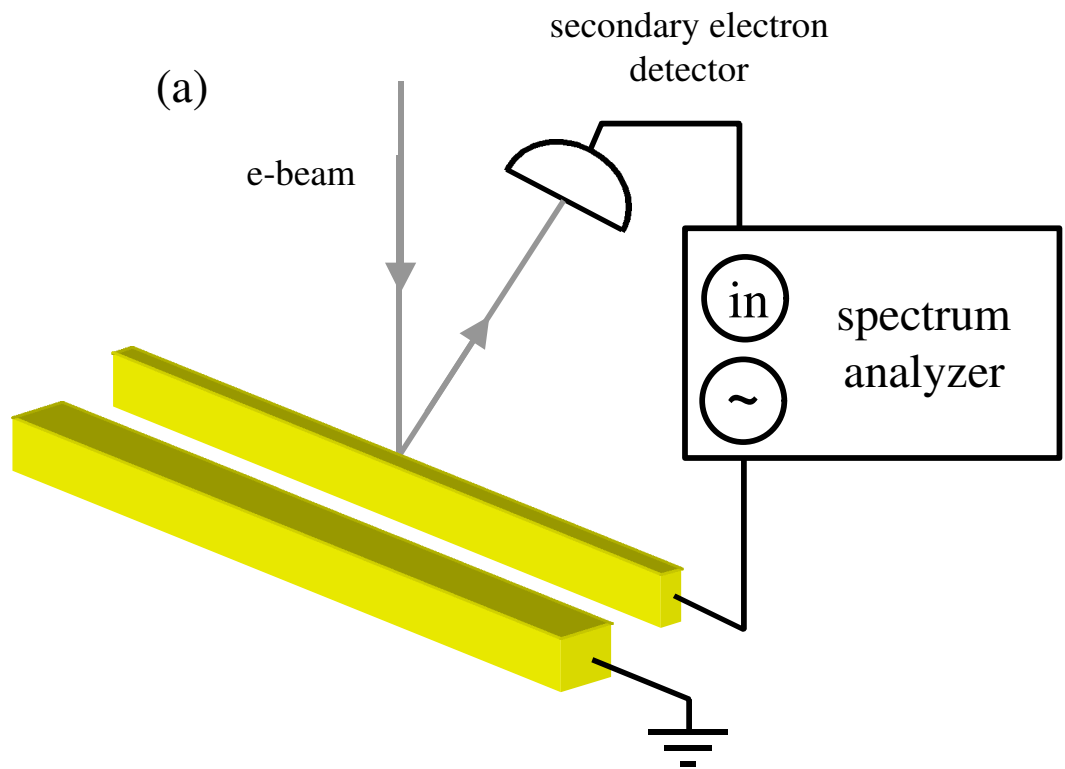
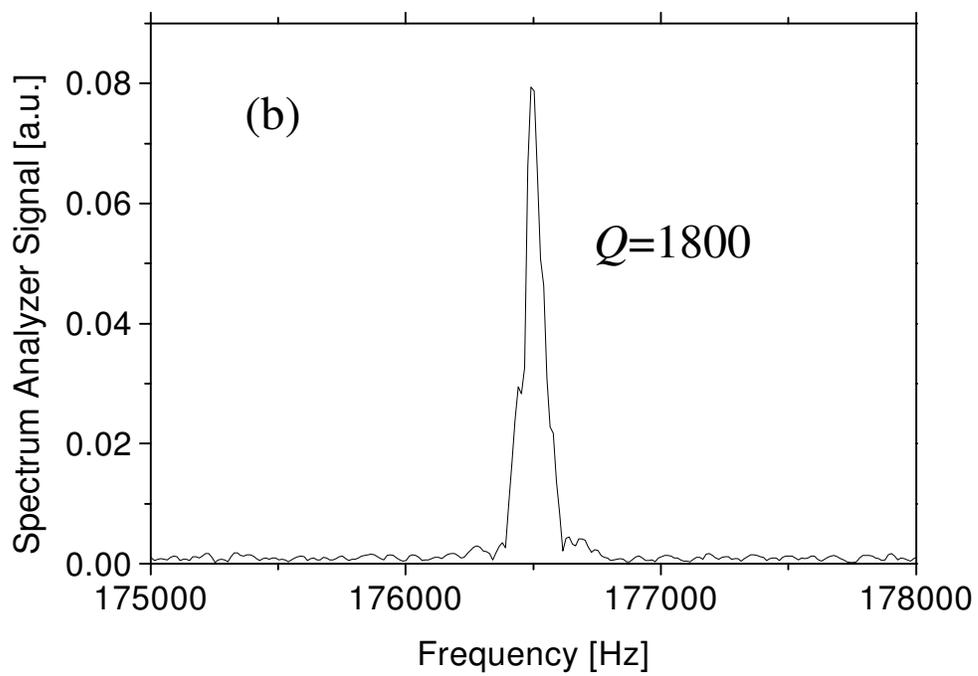
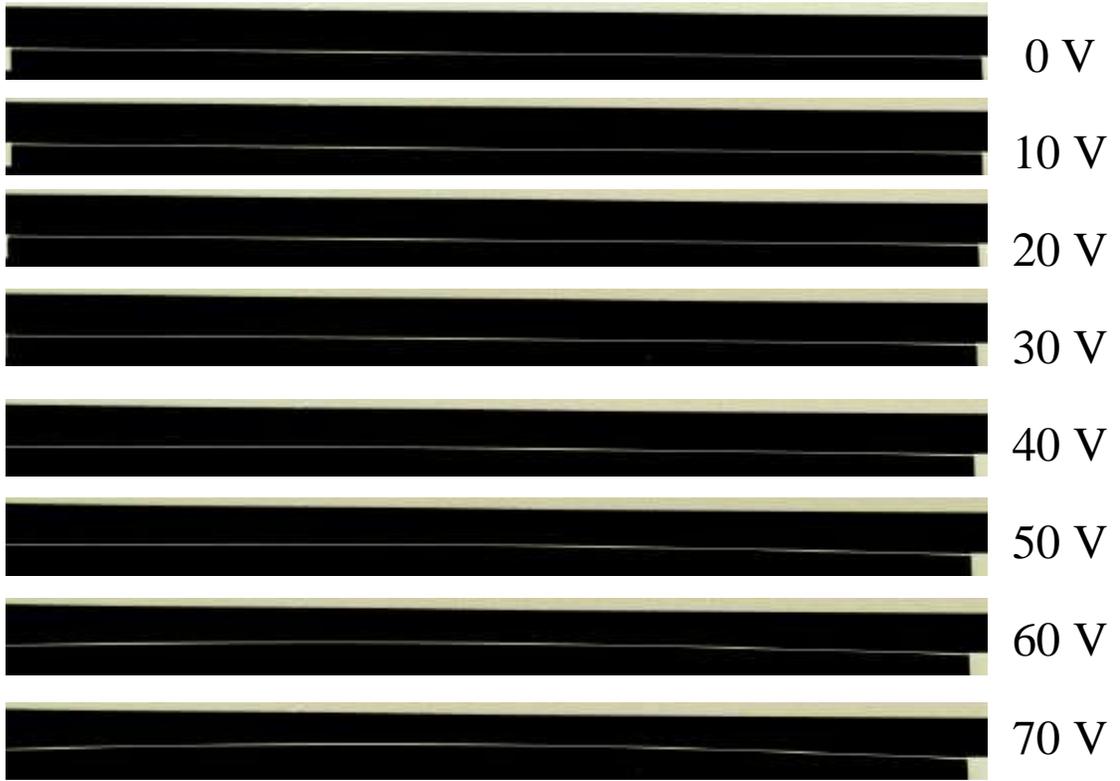
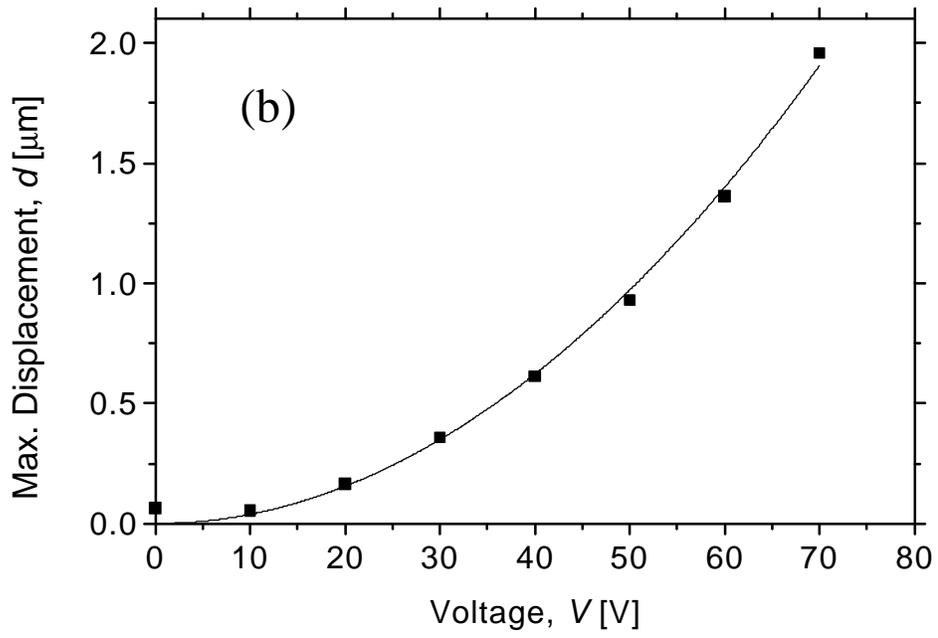
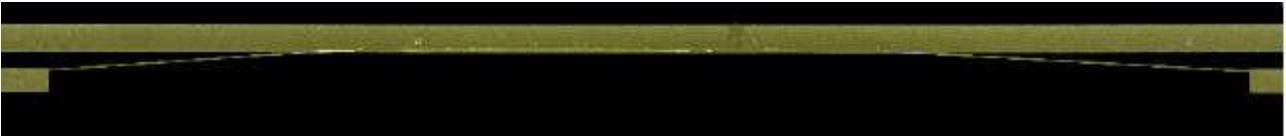
\end{document}